\documentclass[aps,prr,twocolumn,floatfix,amssymb]{revtex4-1} 
\usepackage[utf8]{inputenc}
\usepackage[english]{babel}
\usepackage{amsmath}
\usepackage{amsfonts}
\usepackage{amssymb}
\usepackage{graphicx}
\usepackage{float}
\usepackage{color}
\usepackage{braket}
\usepackage{sidecap}
\usepackage{mciteplus}
\usepackage{xcolor}
\usepackage{sidecap}  
\usepackage[breaklinks=true,colorlinks,citecolor=blue,linkcolor=blue,urlcolor=blue]{hyperref}
\def\Xint#1{\mathchoice
   {\XXint\displaystyle\textstyle{#1}}%
   {\XXint\textstyle\scriptstyle{#1}}%
   {\XXint\scriptstyle\scriptscriptstyle{#1}}%
   {\XXint\scriptscriptstyle\scriptscriptstyle{#1}}%
   \!\int}
\def\XXint#1#2#3{{\setbox0=\hbox{$#1{#2#3}{\int}$}
     \vcenter{\hbox{$#2#3$}}\kern-.5\wd0}}

\def\dashint{\Xint-}
\newcommand{\be}{\begin{equation}}
\newcommand{\ee}{\end{equation}}
\newcommand{\ber}{\begin{eqnarray}}
\newcommand{\eer}{\end{eqnarray}}

\newcommand{\rv}{{\bf r}}
\newcommand{\qv}{{\bf q}}

\def\eer{\end{eqnarray}}

\def\rv{{\bf r}}

\def\qv{{\bf q}}

\begin{document}

\title{Electrically induced charge-density waves in a two-dimensional electron channel: Beyond the Local Density Approximation}
\author{Erica Hroblak, Mohammad Zarenia, and Giovanni Vignale }
\affiliation{Department of Physics and Astronomy, University of Missouri, Columbia, Missouri 65211, USA}

\begin{abstract}
In a previous paper we suggested  that a macroscopic force field applied across a two-dimensional electron gas channel could induce a microscopic charge density wave as soon as the proper compressibility becomes negative, which happens at densities much higher than the critical density for the Wigner crystal transition. The suggestion was based  on a calculation of the ground state energy in the local density approximation.  In this paper we refine our calculation of the energy by including a self-consistent gradient correction to the kinetic energy.  Due to the increased energy cost of rapid density variations, we find a much lower critical density for the onset of the charge density wave.  This critical density coincides with the result of a linear stability analysis of the uniform ground state {\it in the absence of the electric field}.

\end{abstract}

\maketitle
\section{Introduction}
Spontaneous breaking of translational symmetry in electron liquids is a classic problem in condensed matter physics.  The localization of electrons at sufficiently low density was first proposed by Wigner in 1934. \cite{Wigner,Wigner2} He noticed that in this regime the kinetic energy would play a secondary role, allowing the electrons to crystalize in a form that he termed  ``an inverted alkali metal".    A few decades later Overhauser showed that even at the high densities of (real) alkali metals,  exchange and correlation effects would encourage the formation of a charge density wave state (CDW) with wave vector $q \sim 2k_F$, where $k_F$ is the Fermi wave vector.\cite{Overhauser} 
Small ionic displacement would cancel out a large part of the Coulombic energy arising from the non-uniform charge distribution, thus stabilizing the CDW ground state.  More recently, broken translational symmetry has been predicted in stripe and bubble phases in electronic systems at high magnetic field, where the magnetic field plays a crucial role in suppressing the kinetic energy of the electrons.\cite{Fogler1,Fogler2} Stripes have  also been predicted to occur  in high-$T_c$ superconductors, of which the copper-oxide superconductors are the most prominent representatives \cite{emery}.

A different route to stabilizing a CDW state at relatively high density was proposed in Ref. \cite{Erica2017}.  The idea was to ``polarize" a two-dimensional electron channel (2DEC)  by applying a uniform in-plane electric field perpendicular to the edges of the channel: current flow is prevented by insulating barriers running along the channel.   Under ``normal" conditions, the electron gas responds to the field by a small rearrangement of its density, resulting in a small accumulation of electrons along one edge and depletion on the other: translational symmetry is not broken on the microscopic scale.   The situation may change dramatically when the bulk density $n$ of the 2DEC falls below the value ($r_s \sim 2.1$) for which the thermodynamic compressibility $\kappa = (\frac{1}{n^2})(\frac{\partial n }{\partial \mu})$ becomes negative. \cite{Eisenstein,GV,Bello,Steffen,Kusminskiy,Skinner}   A negative compressibility simply means that, due to the dominance of exchange and correlation effect, the chemical potential of the charge-neutral electron gas decreases with increasing density.  By itself, this is not a signal of instability, since the density of the electronic system is stabilized by the requirement of macroscopic charge neutrality.  And indeed, in the last few decades, negative electronic compressibility has been experimentally observed in two-dimensional materials through spectroscopy and measurements of quantum capacitance in transition metal dichalcogenide monolayers,\cite{Riley} carbon nanotubes \cite{Ilani,Wu}, oxide interfaces, and quantum heterostructures.\cite{Li, Larentis,  Wu,  Eisenstein}  This effect has also been observed in graphene when placed under a strong magnetic field. \cite{Skinner} 

Our proposal in Ref. \cite{Erica2017} was to combine  the negative compressibility  with a uniform polarizing electric field.  We were surprised to discover that, as soon as the compressibility becomes negative, the equilibrium density in the presence of the field (obtained from a force balance equation which is equivalent to minimizing the ground-state energy in a local density approximation) develops fast oscillations, which break translational symmetry {\it on the microscopic scale}.  This may happen at densities far higher (by orders of magnitude) than the ones usually considered in the context of Wigner crystallization, CDWs etc..

However, there is a problem with this prediction.  It is well known that the local density approximation for the energy is justifiable only in the limit in which the density is slowly varying on the microscopic scale, which is the scale of the local Fermi wavelength $k_F^{-1}$.    This condition was not satisfied in the calculations of Ref. \cite{Erica2017}:  on the contrary - the oscillations of the equilibrium density were found to occur on the scale of the screening length
$\lambda=\frac{4\pi \epsilon}{e^2}\left\vert\frac{\partial \mu}{\partial n}\right\vert$, which vanishes at the onset of the instability (meaning that the onset of the instability cannot be determined in a physically reliable manner) and becomes at most as large as $\sim 2.8 k_F^{-1}$ at the lowest densities.

The local density approximation is particularly dangerous when applied to the kinetic energy  (in which case it is better known as the Thomas-Fermi approximation).  In this paper we continue to use the local density approximation for exchange and correlation, which is generally found to work well in ab-initio calculations for reasons that are quite well understood, but we improve the treatment of the kinetic energy by introducing a ``gradient correction", which severely discourages rapid variations of the density.  The gradient correction for the kinetic energy density $\epsilon_{\rm kin} [n(\rv)]$  (regarded as a functional of the density $n(\rv)$, wher $\rv= (x,y)$ is the position in space) has the form
\be\label{GradientExpansion}
\delta \epsilon_{\rm kin} [n(\rv)]  = C \frac{\hbar^2}{2m}\frac{|\nabla n(\rv)|^2}{n(\rv)}
\ee
where $C$ is a dimensionless constant that can be derived from linear response theory, as detailed in the Appendix.  For small density variations the value $C=1/4$ provides an upper bound to the energy, meaning that by adopting this value of $C$ we almost certainly overestimate the kinetic energy penalty for forming a charge density wave. Indeed, a significantly smaller value of $C$ will be argued below to be physically more accurate. 

After including  the gradient correction term we repeat the minimization of the total energy of the electron gas in the presence of the force field and find that microscopic translational symmetry is  broken at a value of the Wigner-Seitz radius $r_s \simeq 21.7$ that is much larger than the onset of negative compressibility (yet considerably lower than the typical values associated with the onset of Wigner crystallization  in the absence of external fields).
The  rapid variation of the density in our solution is no longer an issue, since its energy cost is taken into account by the gradient correction term.  {Unfortunately, we also find that the  critical density calculated in this manner coincides with the result of the standard linear stability analysis of the uniform 2DEG in the absence of the electric field, i.e., with the critical density of a spontaneous CDW instability driven by exchange-correlation effects in the local density approximation.  Thus the question whether the uniform electric field actually helps the CDW instability to occur at higher density remains unanswered.}

\section{Description of the physical system and force balance equation}
 As in our previous paper, we consider an electron channel of width $L$ in the $x$ direction, with $-L/2<x<L/2$  and infinite in the $y$ direction.  The system is assumed to be translationally invariant in the $y$ direction, so that the two-dimensional electron density $n(x)$ is a function of $x$ only.  This  density is expressed as $n(x)=n_0+\delta n(x)$ where the uniform component  is exactly neutralized by a positively charged background of  density $n_0$ and
 \be\label{ChargeNeutrality}
 \int_{-L/2}^{+L/2} dx~ \delta n(x)=0\,. 
 \ee

From now on we express $x$ in units of $L/2$, i.e. we set $x=\bar x L/2$.  The total energy per unit area of the system is expressed as a functional of the electronic density $n(x)$ as follows
\begin{widetext}
\be\label{EnergyFunctional}
\frac{E[\bar n(\bar x)]}{{\rm Ry}/a^2}=
\frac{1}{2}\int_{-1}^{1}d\bar x \left\{\bar \epsilon(\bar n(\bar x))+\frac{4C}{\bar L^2}\frac{|\bar n'(\bar x)|^2}{\bar n(\bar x)}+\frac{\bar V}{2} \bar x \bar n(\bar x)-\bar L\int_{-1}^{1}d\bar x'\delta \bar n(\bar x)\delta \bar n(\bar x')\ln|\bar x-\bar x'|\right\}\,,
\ee
\end{widetext}
where $Ry = \frac{e^2}{2 a}$ is the Rydberg unit of energy ($\sim 13.6$ eV), $a = \frac{\hbar^2}{me^2}$ is the Bohr radius, $e$ is the absolute value of the electron charge.   We have introduced the following dimensionless variables (all marked by an overbar): (i) the dimensionless 2D density $\bar n = na^2=\frac{1}{\pi r_s^2}$, and similarly $\delta \bar n=(\delta n) a^2$; (ii) the dimensionless energy {\it density} of a  2DEG of uniform electron density $\bar n$: $\bar\epsilon(\bar n)=\frac{\epsilon(\bar n)}{{\rm Ry}/a^2}$; (iii) the dimensionless width of the channel $\bar L=L/a$; (iv) the dimensionless potential energy difference $\bar V=-FL/{\rm Ry}$ generated by a uniform force $F$ across the width of the channel.  $\bar n'(\bar x)$ denotes the derivative of $\bar n(\bar x)$ with respect to $\bar x$. 
The first term in Eq.~(\ref{EnergyFunctional}) is the local density approximation, where $\bar\epsilon$ includes the kinetic energy density and the exchange-correlation energy densities~\cite{GV}.  The latter is accurately known from the QMC work of Attaccalite et al.~\cite{Attaccalite}.  The second term is the gradient correction to the kinetic energy~\cite{Choudhury,Zarenia1,Zarenia2}, discussed in the introduction.  The third term is the potential energy created by the force applied across the channel and the fourth term is the Hartree energy arising from the inter-particle Coulomb repulsion. 


 The energy functional must be minimized with respect to the density deviation $\delta \bar n(\bar x)$, subject to the constraint~(\ref{ChargeNeutrality}).  This gives us the equation
 \be
 \frac{\delta E[\bar n(\bar x)]}{\delta \bar n(\bar x)}=\mu\,,
 \ee
 where  $\frac{\delta E[\bar n(\bar x)]}{\delta \bar n(\bar x)}$  is the functional derivative of the energy functional with respect to $\bar n(\bar x)$, and $\mu$ is a constant Lagrange multiplier required to satisfy the constraint ~(\ref{ChargeNeutrality}).  Alternatively, we can write
 \be
 \frac{d}{d \bar x}\frac{\delta E[\bar n(\bar x)]}{\delta \bar n(\bar x)}=0\,.
 \ee
Carrying out the prescribed operations and keeping only terms linear in $\delta \bar n$ (under the assumption that $\delta\bar n\ll \bar n_0$) we arrive at the integrodifferential equation
 \be\label{FBE}
\bar\epsilon''(\bar n_0)\frac{\delta \bar n'(\bar x)}{2\bar L}-\frac{4C}{\bar n_0}\frac{\delta \bar n'''(\bar x)}{\bar L^3}-\dashint_{-1}^{1}d \bar x'\frac{\delta \bar n(\bar x')}{\bar x-\bar x'} = -\frac{\bar V}{4 \bar L}\,,
 \ee
where $\bar \epsilon''(\bar n_0)$ is the second derivative of the energy density evaluated at the background density.  This second derivative is related to the compressibility $K(n)$ by the well-known formula $\epsilon''(n)=1/(n^2 K(n))$ and  becomes negative (passing through zero)  for $r_s$ greater than about $2.1$.\cite{GV}   In the above equation $\bar x$ is restricted to the interval $|\bar x|<1$ and the integral is evaluated according to the Cauchy principal value prescription (denoted by a strike across the integral sign).  For $C=0$ Eq.~(\ref{FBE})  reduces to the force balance equation used in Ref. \cite{Erica2017}.  Notice that the equation is linear in $\delta \bar n$, so the induced density is directly proportional to the force applied, $F=-V/L$.
 \begin{figure}
\centering
\includegraphics[width=\columnwidth]{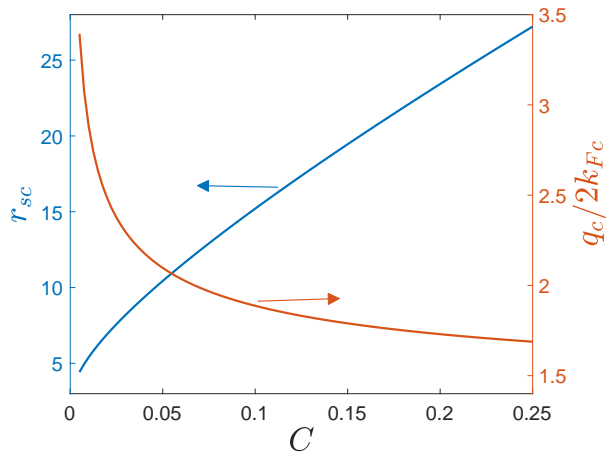}
\caption{Critical Wigner-Seitz radius $r_{sc}$ (blue curve) and the wave vector of the CDW instability  at the critical Wigner-Seitz radius $q_c/(2k_{Fc})$ (red curve) vs coefficient of the gradient correction term $C$.}
\label{figRs}
\end{figure}

 \section{Approximate solution}
 
 In the limit of large channel width (i.e., $\bar L \to \infty$) it may appear that only the Hartree term on the left side of Eq.~(\ref{FBE}) is relevant.  Neglecting the remaining terms would give the elegant classical solution
 \be
 \delta \bar n(\bar x) = -\frac{\bar V}{4 \bar L} \frac{\bar x}{\pi\sqrt{1-\bar x^2}}\,.
 \ee 
 
 This conclusion is correct only as long as the density remains slowly varying on the macroscopic scale ($L$).   In our previous paper we showed that a negative compressibility can induce rapid variations of the density on a microscopic scale ($a$), so that the  derivatives of the density on the left hand side of Eq.~(\ref{FBE}) must be taken into account.  Unfortunately, rapid variation of the density also means that the validity of the local density approximation for the energy (i.e., the first term on the the left hand side of Eq.~(\ref{FBE})) is dubious.  Hence the importance of including the gradient correction term, which strongly discourages such rapid variations by imposing a large energy penalty.
 
 Based on these considerations, we  seek a solution of Eq.~(\ref{FBE})  in the form
 \be \label{Ansatz}
\delta \bar n(\bar x) = -\frac{\bar V}{4 \bar L}\left\{\frac{\bar x}{\pi\sqrt{1-\bar x^2}}+f\sin{(\bar q \bar x)}\right\}\,,
\ee
\\
that is to say the classical solution plus an oscillating term with dimensionless wave vector $\bar q = qL/2 =  (qa) \bar L/2$.  We assume $q>0$, while the real amplitude  $f$ can have either sign as needed.     This Ansatz has been shown in Ref. \cite{Erica2017} to yield excellent results vis-a-vis the exact numerical solution of the integro-differential equation. 

We  substitute Eq.~(\ref{Ansatz}) into Eq.~(\ref{FBE}) and  focus on the region $|\bar x| \ll 1$ (which, we emphasize, can be macroscopically large if  $\bar L$ is large). In this region we have

\be
\dashint_{-1}^{+1} d\bar x' \frac{\sin (\bar q \bar x')}{\bar x-\bar x'} \simeq -\pi \cos (\bar q \bar x) +\pi-2Si(\bar q)\,,~~~|\bar x|\ll 1\,,
\ee
where ${\rm Si}(\bar q)\equiv \int_0^{\bar q}\frac{sin (t)}{t} dt$ is the Sine Integral function and the  strike across the integral sign mandates  that we take the principal value of the integral.   
Thus, we arrive at the following equation for $\bar q$ and $f$:
\begin{widetext}
\be
\left\{\left[\frac{\bar\epsilon''(\bar n_0)}{2\pi}\left(\frac{\bar q}{\bar L}\right)+\frac{4C}{\pi \bar n_0}\left( \frac{\bar q}{\bar L}\right)^3+1 \right]  \cos (\bar q \bar x) - [1-2Si(\bar q)/\pi]\right\} f  +\frac{\bar\epsilon''(\bar n_0)}{2\pi^2 \bar L}=0
\ee
\end{widetext}
(we have omitted a term proportional to $C/\bar L^3$, which is negligible in the limit $\bar L\to \infty$). 

The consistency of our Ansatz requires the coefficient of $\cos (\bar q\bar x)$ to vanish, thus determining $\bar q/\bar L$ as the positive solution of the equation
\be\label{EQ1}
\frac{\bar\epsilon''(\bar n_0)}{2\pi}\left(\frac{\bar q}{\bar L}\right)+\frac{4C}{\pi \bar n_0}\left( \frac{\bar q}{\bar L}\right)^3+1=0\,.
\ee 
Notice that $\bar q/\bar L = qa/2$ so we expect a solution with $qa/2$ on the order $1$, independent of $L$ in the limit $L \to \infty$. This corresponds to  a microscopic charge density wave. 

Given that $C>0$, it is immediately evident that a solution can exist only if the compressibility is negative, i.e. for $\bar\epsilon''(\bar n_0)<0$.   The plain local density approximation $C=0$ yields a solution for any density such that  $\bar\epsilon''(\bar n_0)<0$, but then the wave vector of the solution diverges when the density is close to the the critical value for which $\bar\epsilon''$ vanishes.   On the other hand, a sufficiently large value of $C$ may completely prevent the existence of a solution, by making the left hand side of Eq.~(\ref{EQ1}) always positive for $\bar q>0$. 

For a given value of $C$ a solution of Eq.~(\ref{EQ1}) first appears at a critical Wigner-Seitz ratio $r_{sc}$ and wave vector $\bar q_c$ such that
\be\label{EQ2}
 \frac{\bar\epsilon''(r_{sc})}{2\pi}+ 12 C r_{sc}^2\left( \frac{\bar q_c}{\bar L}\right)^2=0\,,
 \ee 
 i.e., when the zero of the left hand side of Eq.~(\ref{EQ1}), regarded as a function of $\bar q$, yields the minimum of that function.
 Combining Eqs.~(\ref{EQ1}) and~(\ref{EQ2}) we find
 \be\label{QCcrit}
 \frac{\bar q_c}{\bar L} = -\frac{3\pi}{\bar\epsilon''(r_{sc})}\,,~~~~~C=\left(-\frac{\bar\epsilon''(r_{sc})}{6\pi r_{sc}^{2/3}}\right)^3\,.
 \ee
 It is convenient to express the wave vector $q_c$ in terms of the Fermi wave vector $k_{Fc}$ at the critical density.  Noting that $k_{Fc}a= \sqrt{2}/r_{sc}$ we get
 \be\label{Qcrit}
 \frac{q_c}{k_{Fc}}=-\frac{6\pi r_{sc}}{\sqrt{2}\bar\epsilon''(r_{sc})}\,.
 \ee
 
In Fig. \ref{figRs} we plot $q_c/k_{Fc}$ and the critical Wigner-Seitz radius $r_{sc}$ as  functions of $C$. Clearly the plot is meaningful only for $\bar\epsilon''(r_{sc})<0$, i.e., for $r_{sc}>2$.  Numerical values of $\bar\epsilon''(r_{s})$ are taken from Ref.~\cite{Attaccalite}.   We note in passing that an excellent approximation (for the present purpose)  to $\bar\epsilon''(r_{s})$ for $r_s>2$ is 
\be
\bar\epsilon''(r_{s}) \simeq  -1.35 \pi (r_s - 2)\,,~~~~~~r_s>2\,.
\ee

 For a given value of $C$ in the range $0<C<1/4$, where $C=0$ neglects the gradient correction and $C=1/4$ is an upper bound (see next section) the blue curve in Fig. \ref{figRs} allows us to read the value of the critical $r_s$, which is then  used in Eq.~(\ref{Qcrit}) to determine $q_c/(2k_{Fc})$ (red curve).  We see that even in the most unfavorable case, $C=1/4$, our model predicts a charge density wave instability with $q_c \simeq  3.38 k_{Fc}$ and  critical $r_s \simeq 27$, which is smaller than the critical $r_s$ for Wigner crystallization in 2D.   However, there are good reasons to believe that the appropriate value of $C$ is significantly smaller than $1/4$.  Indeed, a self-consistent gradient correction, described in the next section, produces a lower value of $r_{sc}$.
%
%
For completeness, we also calculate the amplitude $f$ of the charge density wave (normalized by $\bar V/(4\bar L)$).  This is  given by 
\be
f =\frac{\bar\epsilon''(\bar n_0)}{2\pi^2 \bar L[1-2Si(\bar q)/\pi]}
\ee
In the limit of large $\bar L$ we make use of the limiting form 
\be
[1-2Si(\bar q)/\pi] \to  \frac{2}{\pi \bar q}\cos\bar q\,,~~~~~\bar q\gg 1\,,
\ee
to arrive at
\be
f =  \frac{\bar\epsilon''(\bar n_0) (qa)}{8 \pi \cos(qL/2)}
\ee
which is finite, but exhibits a wild non-analytic dependence on $L$.  {We will return to this point in the concluding section}.

\section{Self-consistent gradient correction}

In this section we analyze more closely the nature of the gradient correction to the kinetic energy density of a weakly inhomogeneous 2DEG ($\delta n(\rv) \ll n_0$).    To second order in $\delta n$, the kinetic energy density relative to the homogeneous state is given by
\be\label{EE}
\epsilon_{\rm kin}[\delta n]= -\frac{1}{2 A} \int \frac{d^2 q}{(2\pi)^2}\chi_0^{-1}(q)|\delta \tilde n(\qv)|^2\,,
\ee
where $\chi_0(q)$ is the static Lindhard function (i.e., the static density-density response function of the noninteracting 2DEG),  $\delta \tilde n(\qv)$ is the Fourier transform of $\delta n(\rv)$ at wave vector $\qv$, and $A$ is the area of the system. 
The static Lindhard function is known analytically:
\be
\chi_0(q)=-\frac{m}{\pi \hbar^2}\left[1-\Re e \sqrt{1-\frac{4 k_F^2}{q^2}}\right]\,.
\ee

\begin{figure}
\centering
\includegraphics[width=\columnwidth]{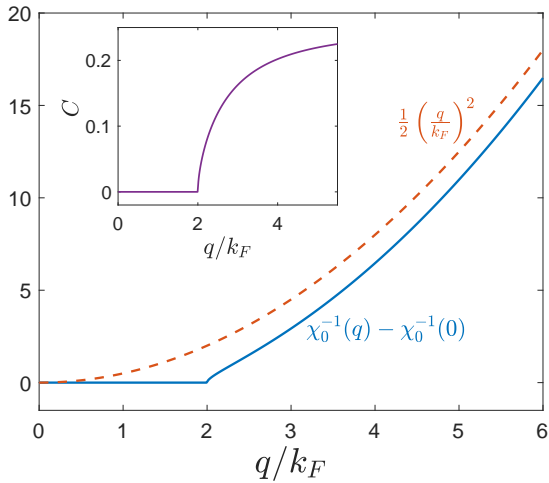}
\caption{Plot of $-\frac{m}{\pi \hbar^2}\left[\chi_0^{-1}(q)-\chi_0^{-1}(0)\right]$ (blue-solid line) and its large-$q$  asymptote $q^2/2k_F^2$  (red-dashed line)  vs $q/k_F$. Inset shows  $C(q)$ from Eq.~(\ref{CC}) vs  $q/k_F$.}
\label{figX}
\end{figure}

It is well-known that the local density approximation for the kinetic energy is obtained (to second in $\delta n$) by ignoring the $q$-dependence of $\chi_0(q)$, i.e., by setting $\chi_0^{-1}(q)=\chi_0^{-1}(0)$ in Eq.~(\ref{EE}).  The exact correction to the local density approximation for the kinetic energy is therefore given by
 \be\label{EEG}
\delta \epsilon_{\rm kin}[\delta n]= -\frac{1}{2 A} \int \frac{d^2 q}{(2\pi)^2}\left[\chi_0^{-1}(q)-\chi_0^{-1}(0)\right]|\delta \tilde n(\qv)|^2\,,
\ee
The  quantity $-\left[\chi_0^{-1}(q)-\chi_0^{-1}(0)\right]$ is plotted in Fig. \ref{figX} (in units of $m/(\pi\hbar^2$).  It is strictly zero for $q<2k_F$ and remains always lower than the parabola $q^2/2k_F^2$, to which it tends for  $q \to \infty$. 
It is easy to verify that the gradient correction of Eq.~(\ref{GradientExpansion}), with $C=1/4$ is obtained precisely by approximating  $-\left[\chi_0^{-1}(q)-\chi_0^{-1}(0)\right]$ by its high-$q$ limit $\frac{\pi \hbar ^2 q^2}{2mk_F^2}$. Therefore Eq.~(\ref{GradientExpansion}), with $C=1/4$, definitely  overestimates  the kinetic energy cost of the inhomogeneity. 
{We emphasize that Eq.~(\ref{GradientExpansion})  with the choice $C=1/4$ provides a rigorous upper bound to the noninteracting kinetic energy functional in the linear response regime.  This fact does not seem to have been recognized in the literature so far.}

From Fig. \ref{figX} we also see that density waves with wave vectors smaller than $2 k_F$ are adequately described by the local density approximation.   Therefore the choice $C=1/4$ is inconsistent, since it predicts a density wave with $q \simeq  3.38 k_{Fc}$  (at $r_s \simeq 27$), which in turn would yield a value of $C$ smaller than 0.2 as can be seen in  the inset of Fig. 2.  However, the choice $C=0$ would also be inconsistent, since it would predict wave vectors much larger than $2k_F$, which would imply $C=1/4$.
These considerations lead us to propose that the value of the coefficient $C$ in the kinetic gradient correction can be chosen consistently with the value of $q_c$ that it predicts. In other words, we set
\be\label{CC}
C(q) = -\frac{m}{\pi\hbar^2}\frac{\chi_0^{-1}(q)-\chi_0^{-1}(0)}{2 (q/k_F)^2}\,,
\ee
such that $C\to 1/4$ for $q \to \infty$ and  $C=0$ for $q<2k_F$. 

{Substituting Eq.~(\ref{CC}) into our Eq.~(\ref{EQ1}) and noting that $\chi_0^{-1}(0)=-\frac{\pi \hbar^2}{m} \bar\epsilon_0''$ where $\bar \epsilon_0''$ is the second derivative of the dimensionless kinetic energy density, we see that Eq.~(\ref{EQ1}) can be rewritten  as 
\be\label{chim1}
\frac{m}{\pi\hbar^2}\chi_0^{-1}(q/k_F)-\frac{\bar\epsilon_{xc}''}{2\pi} -\frac{\sqrt{2} r_s}{q/k_F}=0
\ee
or, after restoring physical units, as
\be
\chi^{-1}(q) \equiv \chi_0^{-1}(q)-\epsilon_{xc}''-\frac{2\pi e^2} {q}=0\,.
\ee
The quantity on the left hand side is recognizable as the inverse of the density-density response function when exchange-correlation effects are treated in the local approximation~\cite{GV} (i.e., when the exact exchange-correlation field created by a density fluctuation $\delta n(\rv)$ is approximated as  $\epsilon_{xc}'' \delta n (\rv)$).  Thus our instability criterion, Eq.~(\ref{EQ1}),  reduces to the requirement that the approximate density-density response function tends to infinity at some value of $q$, signifying an instability with respect to the formation of a charge density wave at that wave vector.  This result is not surprising in the framework of the linear response approximation, which we have adopted here as well as in Ref.~\onlinecite{Erica2017}.  
Plotting the left hand side of Eq.~(\ref{chim1}) vs $q$ for different values of $r_s$ (see Fig. 3) we find that a zero first appears at $r_s\simeq 21.7$, with $q \simeq 3.1 k_F$ and $C =0.17$.  Notice that $q$ and $C$ are no longer exactly given by Eqs.~(\ref{QCcrit}) and (\ref{Qcrit}), since the self-consistent $q$-dependence of $C$ invalidates the derivation of those equations.  
\begin{figure}
\centering
\includegraphics[width=\columnwidth]{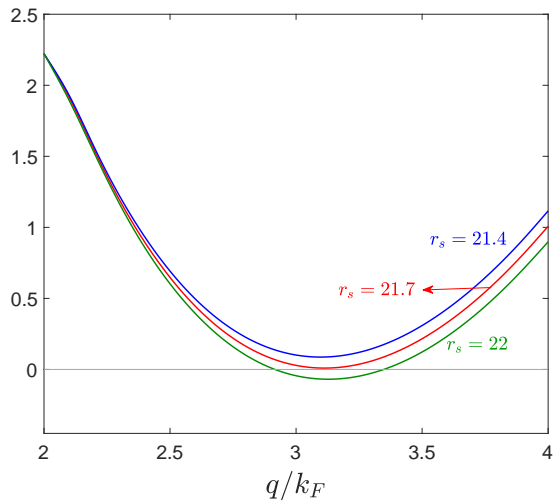}
\caption{Plot of the inverse density-density response function (left hand side of Eq.~(\ref{chim1}), plotted on the vertical axis) vs $q/k_F$ (horizontal axis)  for three different values of $r_s$, from top to bottom,  $r_s=21.4$ (blue),  $r_s=21.7$ (red), and $r_s=22.0$ (green).  The inverse response function first vanishes at $r_s\simeq 21.7$ and $q/k_F \simeq 3.1$.}
\end{figure}




\section{Discussion and outlook}
{The calculations presented in this paper show that the suggestion of  Ref.~\cite{Erica2017}, namely that a macroscopic electric field can induce a microscopic charge density wave in a 2DEG at relatively high density, remains unproven.   
The self-consistent gradient approximation for the kinetic energy density in the linear response regime   turns out to be equivalent to the standard linear stability analysis for the uniform jellium in the absence of an electric field. The critical values of $r_s$ and $q$ are obtained from the ``earliest" zero of the inverse density-density response function $\chi^{-1}(q,r_s)$  {\it in the absence of} an electric field.  This is a spontaneous instability driven by exchange-correlation effects and not attributable to the presence of the macroscopic electric field.  In addition, the amplitude of the charge density wave could not be pinpointed within our linear response approach, which is not surprising given that the appearance of a microscopic CDW in the absence of an external field of the same wave vector is an intrinsically nonlinear phenomenon. 

The way out of these difficulties  is to work out the full numerical solution of the Kohn-Sham equation, which is nonlinear in the electric field, and map the threshold of the CDW instability  as a function of both density and electric field.  This remains a project for the future.}

\section{Acknowledgement}
 We gratefully acknowledge support for this work from  NSF Grant No. DMR-1406568.

\section{References}
%

\end{document}